# Design and Modeling Billing solution to Next Generation Networks

**Kamaljit I. Lakhtaria**
Atmiya Institute of Technology & Science, Rajkot – 360001, Gujarat,
Email:kamaljit.ilakhtaria@gmail.com
**Dr. N. N. Jani**
Dean, Computer Science, Kadi Sarva Vishva Vidyalaya (Deemed University), Gandhinagar, Gujarat
Email: drnnjanicsd@gmail.com

-------------------------------------------------------------------ABSTRACT-------------------------------------------------------------------
Next generation networks (NGN) services are assumed to be a new revenue stream for both network operators and service providers. New services especially focused on a mobile telecommunications that would be used not only as a communication de vice but also as a personal gateway to order or consume a variety of services and products [1]. This type of advanced services can be accomplished when the adaptability of the packet-networks (Internet) and the quality of service of the circuit switched networks are combined into one network [2]. New challenges appear in the billing of this heterogeneous multi services network. Some examples of such a services and possible solutions about charging and billing are examined in this paper. The first steps of mathematical model for billing are also considered.

## 1. INTRODUCTION

Moving to an NGN networks means replacing the existing core networks, which largely dedicated for one service, like PSTN, by a new core network designed to support multi services [2]. Such a network will define a new framework architecture open to different access networks. This will promote a competition and encourage private investments [3]. Charging, billing and payment functions are always important aspects for the success of any business. Since old billing strategies are not enough for the NGN services, therefore new billing models are proposed and a new billing requirements and options are also appears. In general to provide a successful billing system the following points have to be addressed [1]:

- Types of services.
- Charging attributes.
- Timing considerations.
- Payment options.

To understand the new services expected by next generation network, first a general sense about this network structure and components need to be clarified. *Eurescorn-project PI 341* "NGN Services Concepts" has proposed a reference architecture model for NGN. This model will-be discussed by briefly describing the function of its components. Then the traditional billing methods, as well as the new charging attributes, challenges and options are considered.

## 2. ARCHITECTURE OF NGN

Next generation network can be described in general as a packet-based network able to provide a QoS-enabled telecommunication services in which the transport layer is completely separated from the services layer [2,3]. The NGN, also, must include the following features:

- Users of the network have the ability to access different service providers [3].
- It is open to third-party service providers [2].
- Support for wide range of services [3].
- Different access networks exist in the transport layer, but they have to be all independent from the services layer [3].
- Generalized mobility [3].
- Continuity of service; as a result from the last two points the user expected that the service will not interrupted if he/she roam from one access network to another [2].

Figure 1 shows the NGN architecture model that consist of a service creation environment and applications server layer which includes software programs to execute the available services content and interface [2]. As well, the NGN model contains the following network service components:

**Framework**

The framework is a component that is responsible for managing of access-control and security tasks. It also performs additional tasks, e.g., QoS monitoring, NATs and proxies [2,1]. The main function of access-control in the NGN reference architecture abbreviated as AAA, and could be described as following [1]:

- Authentication - the process of retrieving the user identity,
- Authorization - involves checking the identity of the user and the services that user is qualified to access,
- Accounting - collecting the usage information for billing.

Security (e.g. data encryption) is also an important function of framework component. It is required whenever there is a need for confidentiality. It is derived by a number of means; the most common two are hop-by-hop and end-to-end security [1].

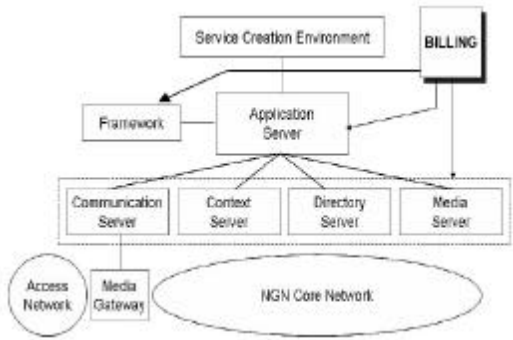

Figure 1: NGN - Reference architecture.

- **Communication server layer:** This layer handles the switching technology that used to support service delivery. Many issues are involved in this layer including the technologies for handling both synchronous and asynchronous communications [2].

- **Context server layer:** Dynamic information about the users and devices are provided by this server, for example location, availability, current session information [2].

- **Directory server layer:** Static information about the users and devices, such as user profiles, are provided by this server [2].

- **Media server layer:** It represents technologies of interactive communications. For example this server may use a speech synthesis application to read email for the user, or a voice recognition application to accept a voice user commands then interpreting and sending these commands to another service application [2].

- **Media gateway:** This component provides technologies to convert from the access network media format (which is analogue in most cases) to a digital packets format, which is a suitable format for the NGN core network [2].

## 3. TRADITIONAL BILLING STRATEGIES

A task of developing a perfect billing system for a heterogeneous network, like the proposed NGN architecture, is not an easy work. In this section, however, the most commonly known charging strategies that have been used in telecommunication industry are considered. Then the qualifications of these strategies for NGN architecture are discussed.

### 3.1 Flat rate billing strategy

This is normally a fixed tariff paid for a predefined period, for example one month. It is clear that it is a straightforward method and easy to implement by the system. It is also easy to be will understood by the costumer, but it is not best suited for all subscribers, since they are all would pay the same value independent of the amount they have utilizes the service [5].

### 3.2 Duration-Based Billing Strategy

When the user request a service, a new session is initialized (normally this is accomplished by reserving a dedicated channel for the user), then as soon as there is no need for the service, the user must end this session. The system will bill the subscriber according to the time length of service delivery. This scheme is also easy to implement and understood by most of the users, because of its wide usage in Public Switched Telephone Network (PSTN). Its main disadvantage is that once the session is initialized the system will account for all the capacity assigned to the reserved channel regardless of the gaps of unused bandwidth [5].

### 3.3 Volume-Based Billing Strategy

In this strategy, the user pays for the amount of data he/she used. Theoretically with this method the user will pay only for the used bandwidth. In practice, to determine the amount of that bandwidth, there are two alternative approaches. The first is to count the number of packets that have been sent and received, but this may not give the exact value, because of packet size variation. The second is to count the number of bits or bytes that have been sent and received. The disadvantage of this approach, it accounts the headers and a retransmitted data with the total volume [5].

## 4. BILLING FOR CONTENT - THE NEW BILLING MODEL

This strategy based also on the content of the data delivered in charging the user rather than the charging by volume or time only. For example if the user downloads a video file, the charge may be depends on the content of that video. This billing strategy introduces new kind of revenues to the service providers and considered as the most important challenges in the NGN billing systems [6]. Technically this new billing model can be implemented only if an application service is capable to separate the presentation (interface) of the service from its content (data).

This separation ensures that the content of the service is protected while its presentation is been accessible to the user. Because of its simplicity and direct support of presentation/content separation, XML and its subset languages are the most important candidate technologies for this kind of applications [6].

Figure-2 illustrates the XML separation of presentation/content of the service, where XML data file does not contain any presentation information while all the presentation information is contained in the XML style sheet. With this structure the same data file could be presented using different style sheets; each may have different access rights. The procedure of content based billing model passes through three phases which can be explained as follows [6]:

**Phase-I: User Identification**

In this phase, each user must have a credential (set of properties) specifying him according to the service required. Now if the user is identified, he/she will be authorized in terms of the pre given credentials.

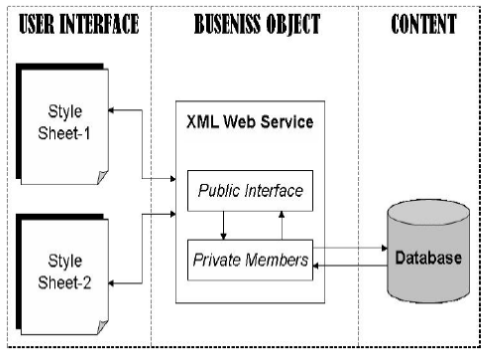

Figure 2: General structure of XML application.

**Phase-II: Attempt to access a resource**

The application will determine the user's appropriate style sheet depending on his credentials (in other words, what he/she is authorized to view from the requested service). For example if a non-member user visits an online journal, in this situation, the user must be authorized only to view the abstracts of the papers (which is normally the case).

**Phase-IIl: Access resolution**

This phase considers the financial transactions that are involved in this model. The system negotiates a service contract with the user. If the user accepts the offer, the financial transaction of the payment is fulfilled.

## 5. OTHER PARAMETERS REQUIRED FOR NGN BILLING SYSTEM

In addition to the billing strategies that have been described in the previous sections, there are some parameters that play an important role in services charging in NGN billing systems. Next some of these parameters are considered.

### 5.1 Quality of Service (QoS)

The QoS is an important issue in the emerging new packet switched network technologies, especially those transporting a real time and interactive applications. The QoS is basically defined as a service agreement between a network user and a network operator "carrier". Which normally represented as service contract contains a set of the following QoS parameters:

1. Peak bandwidth (the maximum allowed bit rate) [1].
2. Average or long-term bandwidth [1].
3. Minimum bandwidth (the minimum acceptable bit rate) [7].
4. Delay (The maximum inter-arrival time between packets) [1].
5. Data delay variation tolerance *"Jitter"* (the maximum acceptable delay variation between packet inter-arrival times) [7].
6. Reliability (It may be illustrated by the percent of packet losses and the received error free packets).

The network services may be differentiated according to the required QoS parameters. For example; voice traffic must be delivered with almost jitter free and the retransmission is almost unacceptable, however, it could accept low reliability in packets delivery. In contrast file transfer traffic could accept higher transfer delay but with higher reliability in packets delivery [1]. Therefore, new generations of data networks must adopt a QoS based billing system. The new billing system must charge users not only for the services, but also for the QoS offered; in other hand the billing system should monitor the network for the QoS service delivery and reacts when the QoS contract is violated from any party [1].

### 5.2 Location

The user location may become an important parameter on NGN billing system. Serving a user at, for example, city centers where the traffic load is high, the operator may charge a higher rate for those certain locations [1].

### 5.3 Network

NGN should facilitate user roaming from one access network to another according to the general mobility feature of the NGN structure. Therefore, not all access networks could provide the same services and QoS parameters. Thus different charges should be applied to different networks according to the offered service and QoS [1].

## 6 NGN SERVICES AND BILLING REQUIREMENTS

The most important feature of NGN is the ability to transport voice, data and video services over one seamless IP-based infrastructure. In general, datagram networks still unable to provide a quality of service similar to which offered by circuit switched networks. However, NGN is faced with two important challenges which are the capability to offer traditional voice call services with similar or better quality that is offered by PSTN based networks, as well as offering new attractive IP services. These challenges must be addressed before migrating to NGN [2].

The traditional business models (voice based ravenous) do not cover the cost of NGN's infrastructure. Therefore, new business models should be introduced to provide IP-based services. This is also an important issue that must be addressed before migrating to NGN [2].

The following examples are some of new services that NGN should offer. A charging scheme is proposed for each one [1]:

- *Voice calls* - These may still be paid by duration-based model.
- *Messaging* **(like e-mails, SMS and MMS)** - A volume-based charging model may be suitable for such services.
- *Videoconferencing and Video telephony* - since these services require high bandwidth and QoS provision, therefore, a duration charging model seems to be suitable for these services.
- *Gaming* - This service is normally an interactive application where a user may select to play with another one. A flat rate billing may works good with this service.
- *Information retrieval* **(for example; internet surfing and use of search engines)** - A subscription and usage-

charging model could be used.

- **Streaming services** - The charging model of this service should consider many parameters including the content, the bandwidth required and the required quality of service. Therefore, charging this service should be based on content and/or duration.
- **Downloading** - Using a content based charging model is recommend for this service.
- **Speech-Recognition or Text-to-Speech** - Usually these services are connected with other services, such as banking and security. Therefore a combination model of a subscription based and application usage based may be better selection (authors' proposal).

### 6.1 Bundling Options

NGN users expect different available services from the network. They will be offered to select the services that meet there pre-references and just pay only for those selection. Also they expect bundling and discounting options on the whole selected services. The usage amount is often bundled into today's networks. Therefore, the NGN must be capable to provide similar bundling option not only for the used minutes, but also for other new parameters like volume and QoS [1].

Network operators should allow third party deals to create new revenues. For example, a special discount on the subscriber device to access a network if the user is buying goods from a particular supermarket, or a special price is offered to costumer who is buying goods using a certain access network. To promote network usage, a volume discount may be offered. For example the user has to pay a certain rate for the first *n* Megabytes, and then this rate is decreased as the volume of data is increased [1].

### 6.2 Taxes

Taxing is one of the most challenging aspects in the billing systems for next generation networks. Traditionally, taxing was carried out by service providers (SP), by adding the required tax amount to the subscriber's bill, and then later the SP pays back the charged amount to the taxation authorities. However, for NGN billing system, the subscribers are capable to roam between different service providers whom might be located in different countries with different taxing legislations and which also might involve a third parity in the billing process. Therefore a taxation in the NGN environment must be addressed and dealing with it carefully [1].

## 7. MATHEMATICAL MODELING OF THE PROBLEM

In order to get a simplified general model for NGN billing process, a study of the traditional billing process is seemed to be helpful. In traditional billing systems, the billing value (charge) (C) can be expressed in the following form:

$$C = f(Q, u, T, P) \quad (7.1)$$

where $Q$ is quantity of data to be transported, $u$ is the pricing unit, $T$ is the duration of the connection and $P$ is the operator policy. The unit price $u$ is also a function of the data rate $R$, the quantity $Q$, the duration $T$, the distance $d$ and the strategy (policy) of the operator $P$. Therefore Equation (7.1) can be written as follows:

$$C = f(Q, u, T, P) \quad (7.2.a)$$

with

$$u = g(R, Q, T, d, P) \quad (7.2.b)$$

Traditional communication systems have two separate general switching technologies; circuit switching and packet switching. In case of circuit switching, that is highly dedicated for telephony services:

- $R$ is fixed, so it has no effect in Equation (7.2).
- Also $Q$ becomes meaningless parameter in the billing model when $R$ is fixed. So, Equation (7.2) becomes:

$$C = f(u, T, P) \quad (7.3.a)$$

with

$$u = g(T, d, P) \quad (7.3.b)$$

The depending of the charge on policy can be put in the following piecewise form:

$$c = \begin{cases} f_1(u_1, T); & P = P_1 \\ f_2(u_2, T); & P = P_2 \\ \vdots \\ f_n(u_n, T); & P = P_n \end{cases} \quad (7.4.a)$$

with

$$u_i = g_i(T, d) \quad (7.4.b)$$

Where $n$ is the number of policies of the operator. As an example, assume that some operator has the following charging policies:

1. Fixed charging unit per distance per time – $P_1$.
2. Flat rate time for period $t$ (between $h_1$ & $h_2$) – $P_2$
3. For the 1st strategy:
   $$C = u(T,d)T$$
   and
   $$u(T,d) = u_o d$$
   Therefore:
   $$C = u_o dT;$$

Where $u_o$ is a constant unit price.

The 2nd strategy has fixed unit price $U_o$ for a given time period. The overall charging system of such an operator may be formed as:

$$c = \begin{cases} u_0 dT & P = P_1 \\ U_0 & h_1 \leq t \leq h_2 \quad P = P_1 \end{cases} \quad (7.5)$$

In case packet switched mode:

- $R$ is represented as bps (bandwidth).

- $Q$ is considered in terms of data volume (Bytes).

So, Equation (7.2) becomes:

$$c = \begin{cases} f_1(Q,u_1,T) & P=P_1 \\ f_2(Q,u_2,T) & P=P_2 \\ \vdots \\ f_n(Q,u_n,T) & P=P_n \end{cases} \quad (7.6a)$$

with

$$u_i = g_i(R,Q,T,d) \quad 1 \leq i \leq n \quad (7.6.b)$$

Now if the service needed is performed through more than one operator, the total charge value can be computed as:

$$C_r = \sum_{k=1}^{K} f_k(Q,u_k,T), \quad P=P_k \quad (7.7a)$$

$$u_k = g_k(R,Q,T,d) \quad 1 \leq i \leq n \quad (7.7.b)$$

Where, $K$ is the number of operators. In this equation the charging policy depends on the agreement between the operator $(i)$ and the operator $(i+1)$. As an example, assume that for a telephone service all the operators are agreed on charging policy of unit price per distance per time, then the total charge $C_T$ becomes:

$$C_T = T \sum_{k=1}^{K} u_k d_k \quad (7.8)$$

Considering the case of NGN, two additional new parameters affect the charging value of service. These parameters are the service content $C_S$ and the bundling options $O_B$. Therefore the charging value $C_{NGN}$ for next generation network can be represented by the following form:

$$C_{NGN} = f_1(Q,u,T,C_S,O_B,P) \quad (7.9)$$

Here also, the unit price $u$ still depends on the data rate, quantity, duration, distance and operator's policy. It will also depend on some new parameters, which are; the quality of service $QoS$, the location $(l)$, the network $(N)$, the content $(C_S)$ and the bundling options $(O_B)$, i.e.

$$u = f_2(R,Q,T,d,QoS,l,N,C_S,O_B,P) \quad (7.10)$$

Finally, the charging policy is also a function depends on the type of service $(S)$, the timing considerations $(t)$, the payment options $(O_P)$ and the marginal profit expected (G), i.e.

$$P = f_3(S,t,O_P,G) \quad (7.11)$$

## 8. CONCLUSION

This paper is part of ongoing work addressing billing aspects for the next generation networks. The evolving of NGN must be based on solid architecture of advanced OSS strategy that is capable to support the new IP service deliver platform (SDP). The emerging IP services have imposed new challenging requirements, including QoS based billing, IP service bundling (IP triple play services) and taxation. This paper addresses some of these billing issues for next generation network. It identifies those challenging issues and proposes how to integrate the existing billing systems into a single and more flexible billing architecture for NGN.

**Authors Biography**

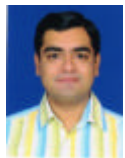

**Mr. Kamaljit I. Lakhtaria** holds the position of Senior lecturer in M.C.A. Programme at Atmiya Institute of Technology & Science, Rajkot. He obtained Master Degree in Computer Application and pursuing research work leading to Ph D beneath the cerebral guidance of Dr. N. N. Jani on "Mobile Communication Technology". He teaches the most advance subjects in MCA and M Sc (IT&CA) since 2005. His inquisitiveness has made him author/present many papers in journals/conferences, seminars, followed by publications at national and international level. The compilation of all his indefatigable efforts has made him author of books.

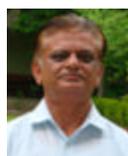

**Dr. N. N. Jani,** has rich experience of 34 years teaching in Computer Science and Applications. At present he is the Professor and Head, Department of Computer Science, Saurashtra University, Rajkot. He has contributed to the first designing syllabus of BCA, MCA, M.Sc.(IT&CA) and updated every two years. He is research guide in Saurashtra University since 1999. Up to the year February 2008, seven-research scholar have completed their Ph.D. from Saurashtra University under his guidance. He is also external guide to Ph.D. programme of North Gujarat University; many scholars have completed their Ph.D. He has written 15 books in the field of Computer Science and Applications, He has worked as peer team member on behalf of NAAC, Banglore.